\documentstyle [pra,aps]{revtex}
\begin{document}
\draft
%
%
\title{A relativistically covariant stochastic model for systems \\ 
with a fluctuating number of particles}
\author{L. M. Morato\thanks{Corresponding author;
e-mail: morato@biotech.sci.univr.it} and
L. Viola\thanks{Present address:
Department of Mechanical Engineering,
Massachusetts Institute of Technology,
77 Massachusetts Avenue, Cambridge, MA 02139;
e-mail: vlorenza@mit.edu} }
\address{ ${}^1$ Facolt\`a di Scienze, Universit\`a di Verona, \\
Strada Le Grazie, 37134 Verona, Italy \\
${}^2$ Dipartimento di Fisica ``G. Galilei'', Universit\`a di Padova, \\
Via F. Marzolo, 8, 35131 Padova, Italy }
\maketitle
%
%
\begin{abstract}
We construct a relativistically covariant stochastic model for systems of
non-interacting spinless particles whose number undergoes random fluctuations.
The model is compared with the canonical quantization of the free scalar field
in the limit of infinite volume.
\end{abstract}
%
%
%
%
%
%
%
\section{Introduction}

Suppose we are given, in the reference frame of an assigned inertial
observer ${\cal O}$, a collection of non-interacting relativistic spinless
particles
with mass $m$ and four-momentum $k:=(\sqrt{ \vec{k}^2 + m^2 }, \vec{k})$
(in units $\hbar$=$c$=1), such that the admissible three-momenta $\vec{k}$
belong to a given finite or countable set $K$.
In this letter we investigate whether it is possible to give a
relativistically covariant statistical description of this simple system in
case the number of particles at fixed $\vec{k}$ is allowed to undergo random
fluctuations.
Actually, the first problem to be addressed is how to model in a Lorentz-invariant
way such a
fluctuation for the single family of particles with given momentum $\vec{k}$.
More in general, this difficulty is related to the longstanding
issue of how to build up a probabilistically consistent description of
stochastic processes within a relativistic setting. Leaving aside a throughout
discussion here, we refer the reader to [1,2] for classical
references on this debated topic and to [3-14]
for more recent contributions, mainly related to stochastic quantization.

In this work, we exploit some ideas introduced within the so-called {\it
comoving approach} to stochastic quantization, as proposed in [12,13].
Roughly speaking, it will be assumed that any observer
$\tilde{\cal O}_k$,
which moves at velocity $\vec{v}_{\vec{k}}:= \vec{k}/\sqrt{\vec{k}^2 + m^2}$
with respect to ${\cal O}$, experiences a fluctuating number of particles $N_k$,
and that such a number can be described through a stochastic process
mathematically well defined on a probability space assigned by $\tilde{\cal O}_k$.

For any fixed momentum $\vec{k}$, the plane waves associated to each
individual quantum particle in the collection
can differ only by a constant phase, which is assumed to be completely
random in $[0, 2\pi)$. Accordingly, the superposition yields a plane wave
whose squared amplitude is related to $N_k$ and whose spatial momentum
is equal to zero in the reference frame of $\tilde{\cal O}_k$.

The key observations are that the application of the Lorentz boost
$\Lambda(\vec{v}_{\vec{k}})$
to this zero-momentum plane wave with fluctuating amplitude provides
a {\it stochastic field} on the
Minkowski space ${\sf M}^4$ and that such a field is {\it Lorentz-invariant by
construction}. The iteration of this procedure for any $\vec{k} \in K$
generates a countable collection of invariant stochastic fields indexed by
$\vec{k} \in K$.

Finally, we examine the case whereby the set $K$ is defined by the allowed
momenta of a free relativistic quantum particle with wavefunction subjected to
periodic boundary conditions in a cubic box introduced in the frame of
${\cal O}$. Performing, as usual, the limit to
infinite volume of the basic statistical invariant relating the stochastic
field at two
different events in ${\sf M}^4$, we find that it is possible to assign the
statistics of the original comoving processes so that the limit coincides
with the two-points Wightman function of the quantized free field.

\section{Construction of the Lorentz invariant stochastic field}

Let ${\cal O}$ be an assigned inertial observer and let
$k:=(\sqrt{ \vec{k}^2 + m^2 }, \vec{k})$ denote the four-momentum of a
zero-charge spinless quantum particle of mass $m$ in the associated
frame of reference (in units $\hbar$=$c$=1). The state of the particle is
represented by the plane wave
\begin{equation}
\phi_k(x)=\mbox{e}^{i (k_\mu x^\mu + \varepsilon)}\;, \hspace{1cm}
x \in {\sf M}^4\;, \label{planewave}
\end{equation}
where $x^\mu=(t,\vec{x})$ are the space-time coordinates associated to
${\cal O}$ and
$\varepsilon \in [0,2\pi)$ is a constant phase factor. To the vector $\vec{k}$
one can associate the constant velocity field
\begin{equation}
\vec{v}_k(\vec{x})=\vec{v}_{\vec{k}}:=
{\vec{k} \over \sqrt{\vec{k}^2+m^2} }\;,
\hspace{1cm}\vec{k} \in K, \vec{x} \in {\sf R}^3\;.
\label{velocity}
\end{equation}
We now introduce an observer $\tilde{\cal O}_k$ which moves with velocity
$\vec{v}_{\vec{k}}$ with respect to ${\cal O}$. We will call this the
{\it k-comoving observer}. (We assume the coordinates of ${\cal O}$ and $\tilde{\cal O}_k$
 coincide at $t=0$)

It can be easily verified that the Lorentz boost associated to the velocity
$\vec{v}_{\vec{k}}$,
$\Lambda (\vec{v}_{\vec{k}})$, transforms the state represented by $\phi_k(x)$
into the zero-momentum plane wave
\begin{equation}
\tilde{\phi}_k(\tilde{t}_k)=\mbox{e}^{-i(m \tilde{t}_k + \varepsilon) } \;,
\hspace{1cm}\tilde{t}_k \in {\sf R}\;,
\label{planewave2}
\end{equation}
where $\tilde{t}_k$ denotes the proper time of the observer $\tilde{\cal O}_k$,
i.e.
\begin{equation}
\tilde{t}_k=\tilde{t}_k(x)=-{1 \over m} \left( \vec{k} \cdot \vec{x} -
\sqrt{ \vec{k}^2 + m^2 }\, t \right)\;, \hspace{1cm}
x \in {\sf M}^4 \;.
\label{comovingtime}
\end{equation}
We stress that the transformed wave function $\tilde{\phi}_k$ does not
depend on the $k$-comoving spatial coordinates.

Let now
\begin{equation}
\phi_k^{(\varepsilon_i)}:= \mbox{e}^{i(k_\mu x^\mu + \varepsilon_i)}\;,
\hspace{1cm}i=1,\ldots,N_k\;,
\label{planewave3}
\end{equation}
denote the plane wave associated to a generic single $i$-th particle in the
collection. We assume that $\{\varepsilon_i \}_{i=1,\ldots,N_k}$ is generated
by a completely random sampling in the interval $[0, 2\pi)$.
The resulting wave function is then
\begin{equation}
\sum_{i=1}^{N_k} \phi_k^{(\varepsilon_i)} = \mu_k \,
\mbox{e}^{i k_\mu x^\mu } \;, \hspace{1cm} x \in {\sf M}^4 \;,
\label{sum}
\end{equation}
where
\begin{equation}
\mu_k  := \sum_{j=1}^{N_k}
\mbox{e}^{i \varepsilon_j} \;.
\label{mu}
\end{equation}
We get then
\begin{equation}
\mu_k \mu_k^\ast = N_k + \sum_{ i,j\not=i =1}^{N_k}
(\cos \varepsilon_i \cos \varepsilon_j + \sin \varepsilon_i \sin \varepsilon_j ) \;.
\label{sum2}
\end{equation}
Under the random sampling assumption made above we finally obtain
\begin{equation}
\lim_{N_k \uparrow \infty} \mu_k \mu_k^\ast = N_k \;.
\label{ennek}
\end{equation}
Alternatively, we can allow $N_k$ to take values in $\left\{ 0,1,2,\ldots\right\} $
and interpret ${\left\{ \varepsilon_i \right\} }_{i = 1, \ldots, N_k}$ as a
set of random variables
which are independent and identically distributed with common uniform
density on $[0, 2\pi)$.
Denoting by $E_{(phases)}$ the average over all possible values of
${\left\{ \varepsilon_i \right\}}_{i = 1, \ldots, N_k}$ we immediately get
\begin{equation}
E_{(phases)} \{\mu_k \mu_k^\ast\} = N_k \;.
\end{equation}

In the reference frame of ${\cal O}$, we associate to the $N_k$-particles
system the field
\begin{equation}
\phi_k^{(N_k)}(x) = \mu_k \, \mbox{e}^{-i k_\mu x^\mu}\;, \hspace{1cm} x \in
{\sf M}^4\;,
\label{field}
\end{equation}
which becomes the real-variable function
\begin{equation}
\tilde{\phi}_k^{(N_k)}(\tilde{t}_k) = \mu_k \,
\mbox{e}^{-i m \tilde{t}_k }\;, \hspace{1cm} \tilde{t}_k \in {\sf R}\;,
\label{function}
\end{equation}
in the comoving reference frame of $\tilde{\cal O}_k$.

Let us now also put
\begin{equation}
\mu_k = \mu^0_k \, \mbox{e}^{i \varphi} \;.
\label{phase}
\end{equation}
Eliminating the overall constant phase $\varphi$ we introduce then the
following complex scalar function in the comoving coordinates:
\begin{equation}
\tilde{X}_k(\tilde{t}_k) = \mu^0_k \, \mbox{e}^{-i m \tilde{t}_k}\;,
\hspace{1cm} \tilde{t}_k \in {\sf R}\;.
\label{ics}
\end{equation}
Bearing in mind that a fluctuation of $N_k$ implies a fluctuation of $\mu^0_k$, we
promote the latter to a stochastic process, denoted by
$\left\{ \tilde{X}^0_k(s), s \in {\sf R}\right\} $:
\begin{equation}
\tilde{X}_k(\tilde{t}_k)= \mu^0_k \, \mbox{e}^{-i m \tilde{t}_k}
\longrightarrow \tilde{X}_k(\tilde{t}_k):=\tilde{X}_k^0
(\tilde{t}_k) \,\mbox{e}^{-im \tilde{t}_k}\;, \hspace{1cm}\tilde{t}_k
\in {\sf R} \;.
\label{randomamp}
\end{equation}
Notice that the process $\tilde{X}_k^0$ must be understood as defined on a
suitable probability space assigned by the $k$-comoving observer
$\tilde{\cal O}_k$.

The image of the process $\tilde{X}_k(\tilde{t}_k)$ in the reference frame of
${\cal O}$ is found by application of $\Lambda(\vec{v}_{\vec{k}})$, namely
\begin{equation}
X_k(x)=  \tilde{X}^0_k(\tilde{t}_k(x)) \, \mbox{e}^{i k_\mu x^\mu} \;,
\hspace{1cm}x \in {\sf M}^4\;,
\label{image}
\end{equation}
with $\tilde{t}_k(x)$ given by (\ref{comovingtime}).
{\it Thus, the relativistic setting forces the (complex) stochastic process
$\tilde{X}_k$ to be transformed into a (complex) stochastic field which is
Lorentz-invariant by construction.}

Notice that for $N_k$ sufficiently large we have
\begin{equation}
X_k(x) X_k(x)^\ast = \tilde{X}^0 _k(\tilde{t}_k(x))^2
\approx \mbox{ \begin{tabular}{l}
       no. of particles with momentum $(m,\vec{0})$  observed \\
       by the $k$-comoving observer at time $\tilde{t}_k(x)\;.$
\end{tabular} }
\label{number}
\end{equation}
Alternatively, we have, for finite $N_k = 0, 1, 2, \ldots $
\begin{equation}
E_{(phases)}\{X_k(x) X_k(x)^\ast\} = E_{(phases)}\{\tilde{X}^0 _k(\tilde{t}_k(x))^2\} =
       \mbox{ \begin{tabular}{l}
       no. of particles with momentum \\
       $(m,\vec{0})$ observed by the $k$-comoving \\
       observer
       at time $\tilde{t}_k(x)\;.$
\end{tabular} }
\label{number2}
\end{equation}

\section{The two-points invariant and a connection with the quantized free
field}

Let now $K$ denote a finite set of three-momenta $\vec{k}$ and let us iterate the
procedure outlined in the previous section for every $\vec{k} \in K$. We
are then left with a collection of Lorentz-invariant stochastic fields that we
represent as
\begin{equation}
\vec{X}(x):=(X_1(x), X_2(x),\dots, X_k(x)), \hspace{1cm}
x \in {\sf M}^4\;,
\label{collection}
\end{equation}
Then
$\vec{X}(x)$ is a stochastic vector field mathematically well defined on a
probability space which
is the product of the individual probability spaces associated to each $k$-comoving
observer $\tilde {\cal O}_k , k\in K$ .

Furthermore, it is possible to introduce in a natural way an invariant scalar
product between complex stochastic vectors which are obtained by considering the field
$\vec{X}(x)$ at two different points in ${\sf M}^4$:
\begin{eqnarray}
( \vec{X}(x_1), \vec{X}(x_2) ) & := & \sum_{\vec{k} \in K}
                E\{ X_k(x_1) X_k^* (x_2)\} \nonumber \\
& = & \sum_{\vec{k} \in K} E\{ \tilde{X}_k^0(\tilde{t}_k(x_1))
\tilde{X}_k^0(\tilde{t}_k(x_2)) \} \mbox{e}^{i k_\mu(x_1^\mu- x_2^\mu) }
\;, \label{scalarpr}
\end{eqnarray}
with $x_1, x_2 \in {\sf M}^4, x_1 \not = x_2$, and the symbol $E$
denoting mathematical expectation over the field realizations.
Both (\ref{collection}) and
(\ref{scalarpr}) are understood to be extended to the countable case in a
formal sense. The function defined by (\ref{scalarpr}) on ${\sf M}^4
\times {\sf M}^4 \rightarrow {\sf R}$ is Lorentz-invariant by construction.

In order to give more explicit expressions, we need to introduce at this point
some physically reasonable assumptions on the family of basic stochastic processes
$\{ \tilde{X}^0_k \}_{\vec{k} \in K}$. Specifically, considering only
the marginal statistics we will assume that there exist real functions
$b:\, K \rightarrow {\sf R}^+$ and $R:\, K \times {\sf R}^+ \rightarrow
{\sf R}$ such that, for every $\vec{k} \in K$,
\begin{eqnarray}
 & i)    & \tilde{X}_k^0(s) \geq 0 \;\;\;\forall s \in {\sf R}\;,
\nonumber \\
 & ii)  &  E\{ \tilde{X}_k^0(s) \}=b(k) < +\infty
 \;\;\; \forall s \in {\sf R}\;, \nonumber \\
 & iii) &  Cov\{ \tilde{X}_k^0(s),\tilde{X}_k^0(s') \}=
  R(k, |s-s'| ) \;\;\; \forall s,s' \in {\sf R}\;,
\label{statistics}
\end{eqnarray}
the symbol $Cov$ denoting as usual the covariance.
Exploiting these assumptions, the two-points invariant function (\ref{scalarpr})
takes the form
\begin{equation}
(\vec{X}(x_1), \vec{X}(x_2) )= \sum_{\vec{k} \in K}
\left\{ R(k, |\tilde{t}_k(x_1)- \tilde{t}_k(x_2)| ) + b(k)^2 \right\}
\mbox{e}^{i k_\mu(x_1^\mu-x_2^\mu)}\;, \hspace{1cm}x_1,x_2 \in {\sf M}^4
\;. \label{choice}
\end{equation}
The simplest rough choice for a stochastic process satisfying (\ref{statistics})
is
\begin{equation}
\tilde{X}^0_k(s) :=
\left\{
  \begin{array}{lc}
    |\nu(s)|          & \mbox{if } |\nu(s)| \leq 2 b(k) \;, \\
    |\nu(s)| - 2 b(k) & \mbox{otherwise} \;,
  \end{array}
\right.
\end{equation}
where $\nu(s)$ denotes a white noise. In this case $\tilde{X}^0_k(s)$ is
independent on $\tilde{X}^0_k(s')$ for all $s' \not = s$ and, therefore,
\begin{equation}
( \vec{X}(x_1), \vec{X}(x_2) ) =  \sum_{\vec{k} \in K}
b(k)^2 \, \mbox{e}^{i k_\mu(x_1^\mu-x_2^\mu)}\;, \hspace{1cm}
x_1,x_2 \in {\sf M}^4\;.
\label{choice2}
\end{equation}

Let now $K$ be defined by the set of the admissible momenta of a quantum
particle of mass $m$ confined to a cubic box of side $L$ in the reference
frame ${\cal O}$ and subjected to periodic boundary conditions.
Exploiting the standard infinite-volume substitution,
\begin{equation}
{1 \over L^3} \sum_{\vec{k}\in K} {\longrightarrow}
{1 \over {(2 \pi)}^3}  \int d^3 k\;,\hspace{1cm}{L \rightarrow \infty}\;,
\label{standard}
\end{equation}
we get
\begin{equation}
{(\vec{X}(x_1),\vec{X}(x_2)) \over  L^3 } \longrightarrow
\int {d^3 k \over {(2 \pi)}^3} \; b(k)^2 \,
\mbox{e}^{i k_\mu (x_1^\mu-x_2^\mu) } \;, \hspace{1cm}x_1,x_2 \in {\sf M}^4\;.
\label{2points}
\end{equation}
It is worth to note that the manifest relativistic invariance is
broken on the left side of (\ref{2points}). However, it can be restored
on the right side by requiring a Lorentz-invariant measure in the momentum space,
i.e. by choosing
\begin{equation}
b(k)^2:= \alpha { m \over  \sqrt{\vec{k}^2 +m^2} } \;,
\end{equation}
$\alpha$ being any positive adimensional constant. Accordingly,
\begin{equation}
{1 \over 2 \alpha m L^3} (\vec{X}(x_1),\vec{X}(x_2)) \longrightarrow
\int {d^3 k \over {(2 \pi)}^3} \; {1 \over 2 \sqrt{\vec{k}^2 +m^2} } \,
\mbox{e}^{i (k_\mu(x_1^\mu-x_2^\mu)} \;, \hspace{1cm} x_1,x_2 \in {\sf M}^4\;,
\label{wightman}
\end{equation}
which can be readily
identified with the familiar two-points Wightman function for the free
relativistic scalar field of canonical quantization [15].

\section{Conclusions}

In this paper we have built up a simple-minded stochastic model for systems
consisting of a fluctuating
number of free, relativistic non-interacting spinless quantum particles,
leading to a Lorentz-invariant stochastic field on the Minkowsky space.

Since no detail concerning the origin of the fluctuations is incorporated in the model,
the set
$K$ of the admissible momenta as well as the statistics of the basic processes
${\left\{ \tilde{X}_k^0 \right\}}_{k \in K}$ are left largely unspecified at this stage.

This fact leaves the possibility of applying the method to different physical
situations.

In particular, defining the set $K$ as the allowed momenta for a free quantum particle
subjected to periodic boundary conditions, it is shown that, under the simplest
assumptions on the marginal statistics of ${\left\{ \tilde{X}_k^0 \right\}}_{k \in K}$,
the arising mathematical structure exhibits some non obvious analogies with the
canonical quantization of the free scalar field.

\acknowledgments
Valuable comments by G. Inverso are gratefully acknowledged.

\end{document}